\documentclass[%
 aip,
 amsmath,amssymb,
 reprint,%
]{revtex4-1}

\usepackage{graphicx}
\usepackage{dcolumn}
\usepackage{bm}
\usepackage[utf8]{inputenc}
\usepackage[T1]{fontenc}
\usepackage{mathptmx}
\usepackage{etoolbox}

\makeatletter
\def\@email#1#2{%
 \endgroup
 \patchcmd{\titleblock@produce}
  {\frontmatter@RRAPformat}
  {\frontmatter@RRAPformat{\produce@RRAP{*#1\href{mailto:#2}{#2}}}\frontmatter@RRAPformat}
  {}{}
}%
\makeatother
\begin{document}

\preprint{AIP/123-QED}

\title[Sample title]{Measurement of surface tension coefficients of liquids based on equal thickness interference}

\author{Xu Ziyi}
\affiliation{ 
Wuhan University, College of Physical Science and Technology, Wuhan, 430072, China
}

\author{Xu Chongyuan}%
\email{moke2001@whu.edu.cn}
\affiliation{ 
Wuhan University, College of Physical Science and Technology, Wuhan, 430072, China
}

\author{Tong Liwen}%
\affiliation{ 
Wuhan University, College of Physical Science and Technology, Wuhan, 430072, China
}

\author{Chen Keyu}%
\affiliation{ 
Wuhan University, College of Physical Science and Technology, Wuhan, 430072, China
}

\author{Dong Ziwei}%
\affiliation{ 
Wuhan University, College of Physical Science and Technology, Wuhan, 430072, China
}

\date{\today}

\begin{abstract}
The surface tension coefficient is a key parameter in fluid mechanics. The conventional method to measure it is to determine the critical surface tension that causes the rupture of a liquid film. However, this method has a large error because the surface tension coefficient is very sensitive to environmental factors. In this paper, we propose a new method based on equal thickness interference to study the mathematical model of the stationary liquid surface shape and measure it by the interference technique. This method enables the accurate measurement of the liquid surface tension coefficient. The experimental results demonstrate that this method has the advantages of high accuracy, low cost, and simple instrumentation. The maximum measurement error is less than 3.1\%. Moreover, this method can be applied to measure the liquid-liquid interfacial tension, which has a good application prospect.
\end{abstract}

\maketitle

\section{\label{sec:level1}Introduction}

The surface tension coefficient of liquids is commonly used in discussing the surface phenomena of liquids and analysing the properties of liquids, among other things. In fields such as civil engineering and hydraulics, parameters such as the liquid surface tension coefficient have an important influence on the collapse flow pattern and kinetic behaviour of a column of wet particles in a pendulum state\cite{artoni2013collapse}. In addition, the "pore size" of a liquid film is also affected by the surface tension coefficient of the target liquid\cite{zhao2013preparation}. Therefore, accurate measurement of liquid surface tension coefficients is of great importance in fundamental physics experiments and engineering applications.

Liquid surface tension measurement methods can be divided into two main groups: contact and non-contact methods\cite{neumann2010applied}. In university physics experiments, the contact method is often used for measurement, such as the pull-off method. However, this method has a low measurement accuracy in the laboratory due to the limited accuracy of experimental instruments and the large influence of human factors. Non-contact measurements are generally made using optical methods. Although this type of method has higher measurement accuracy, it also has the disadvantages of high cost and difficult operation.

In this paper, a mathematical model of liquid surface shape is derived for the correlation between the shape of liquid droplets in a container and the surface tension. The surface tension coefficients of the liquid are fitted by projecting a plane beam vertically onto the surface of the liquid and reflecting it on the surface layer and the glass at the bottom of the container to obtain interference fringes, which are combined with the derived model of liquid surface shape. In addition, the method is extended to the measurement of liquid-liquid interfacial tension. It is experimentally verified that this method is easy to use, low cost, high precision and has a wide range of application prospects.

\section{Experimental principles}

\subsection{Liquid surface shape}

The liquid surface in a cylindrical container has axial symmetry, and it is only necessary to study the shape of the cross-section curve along the axis. As shown in Fig. 1, The coordinate is set, and the origin is the intersection of the bottom of the liquid surface and the $Z$ axis. $P_{00}$ and $P_r$ are pressures at the point $(0,0)$ and the point $(r,z(r))$, both of which are below the liquid surface. The liquid measured in the experiment can saturate the container and form a concave liquid surface.

\begin{figure}
\includegraphics[width=\linewidth]{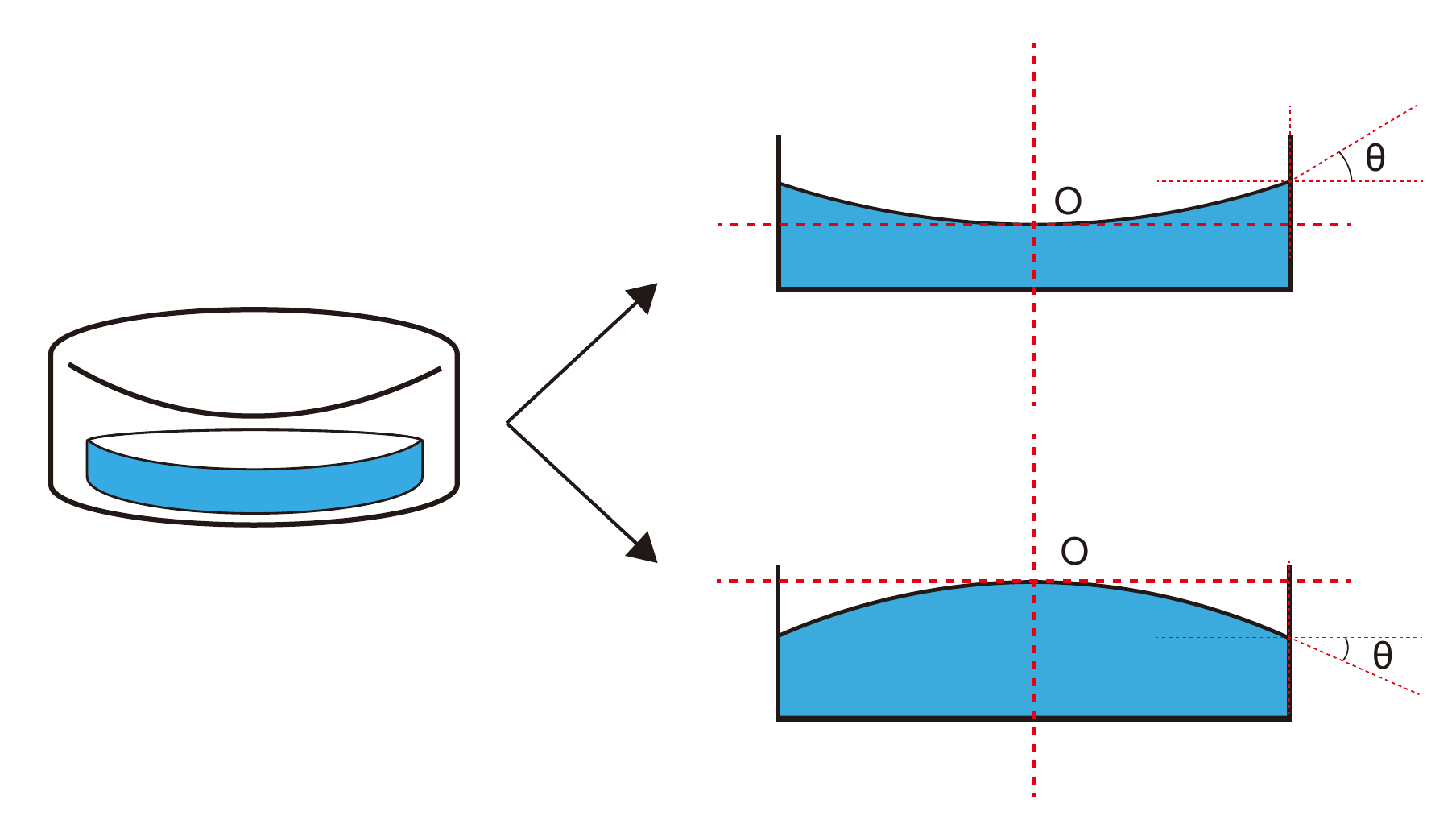}
\caption{\label{fig1} Coordinate system of the axisymmetric liquid surface and force analysis of the liquid surface in two cases.}
\end{figure}

It’s known that the relation of $P_{00}$ and $P_r$ can be given as,

\begin{equation}
P_{00}+\rho gz(r)=P_r    
\end{equation}

where $\rho$ is the density of the liquid.

However, both $P_{00}$ and $P_r$ are hard to measured directly.

According to the Young-Laplace equation\cite{chen2006derivation}, $P_{00}$ and $P_r$ can be expressed as, 

\begin{equation}
\begin{array}{cc}
\begin{cases}
P_{00}=\Delta P_0+P_0 \\
P_{r}=\Delta P_r+P_0
\end{cases}
\end{array}
\end{equation}

where $P_0$ is the standard atmospheric pressure and $\Delta P_0$ and $\Delta P_r$ represent pressure differences across the fluid interface (the exterior pressure minus the interior pressure) at the point $(0,0)$ and the point $(r,z(r))$, respectively, and they are given as,

\begin{equation}
\begin{array}{cc}
\begin{cases}
\Delta P_0=-\frac{2\sigma}{R_0} \\
\Delta P_{r}=-\sigma (\frac{1}{R_1}+\frac{1}{R_2})
\end{cases}
\end{array}
\end{equation}

where $\sigma$ is the surface tension coefficient. Since the liquid surface has axial symmetry and the two principal radii of curvature are equal at the bottom of the liquid surface, let both be $R_0$, $R_1$ and $R_2$ are the two principal radii of curvature of the surface at the point (r,z(r)), and they can be expressed as\cite{ivaki2019deforming}

\begin{equation}
\begin{array}{cc}
\begin{cases}
\frac{1}{R_1}=\frac{z_{rr}}{(1+z_r)^{\frac{1}{2}}}\\
\frac{1}{R_2}=\frac{z_{r}}{r(1+z_r)^{\frac{1}{2}}}
\end{cases}
\end{array}
\end{equation}

where $z_r$ and $z_{rr}$ are the first-order and second-order derivatives of $z$ over $r$. 

Substituting equations (2)-(5) into equation (1), the equation of the cross-section curve of the liquid surface can be obtained as,

\begin{equation}
\sigma\left[\frac{rz_{rr}+z_r}{r(1+z_r)^{\frac{1}{2}}}-\frac{2}{R_0}\right]=\rho gz(r)   
\end{equation}

\subsection{Equal Thickness Interference}

Parallel beam incident perpendicular to the container, the beam is reflected in the surface layer of the liquid and the bottom glass of the container, respectively, outside the liquid surface of the two beams of reflected light will be interfered with reference\cite{fowles1989introduction}, let the wavelength be the refractive index of the liquid is , then the distance from the center of the circle At the liquid surface of the light reflected with the light reflected on the bottom glass optical range difference is:

\begin{equation}
\Delta l=2nz(r)
\end{equation}

When the interfering light is exactly fully intensified or exactly canceled, the required phase difference between the two beams of light is $\pi+2k\pi$ or $2k\pi$, so the height difference between neighboring interfering rings (bright or dark) is $\frac{\lambda}{2n}$. Therefore, by measuring the radius of the bright ring at each level and combining it with the height at which the ring at that level is located, the cross sectional curve of the liquid surface can be obtained as $z(r)$. Under the condition that the density of the liquid is known, the measurement results are fitted with the equation (6). The surface tension coefficient can be obtained by fitting $\sigma$.

\section{Instruments and experiments}

\subsection{Instrument description}

In this experiment, the experimental apparatus mainly includes a reading microscope, a laser source, a beam-expanding mirror set, and a cylindrical container. As shown in Fig. 2, the container is placed horizontally under the reading microscope, and the side in contact with the liquid is smooth plane glass, and the other side is hairy glass.

As shown in fig. 3, in order to facilitate the reading of the interference pattern, a CCD camera was fixed at the eyepiece and connected to the computer in the experiment. A helium-neon laser with a wavelength of $632.8 nm$ \cite{houreld2007irradiation} was selected as the laser source, and a beam expanding mirror was used to enlarge the observation field to avoid damage to the camera caused by an overly strong laser beam.

\begin{figure}
\includegraphics[width=\linewidth]{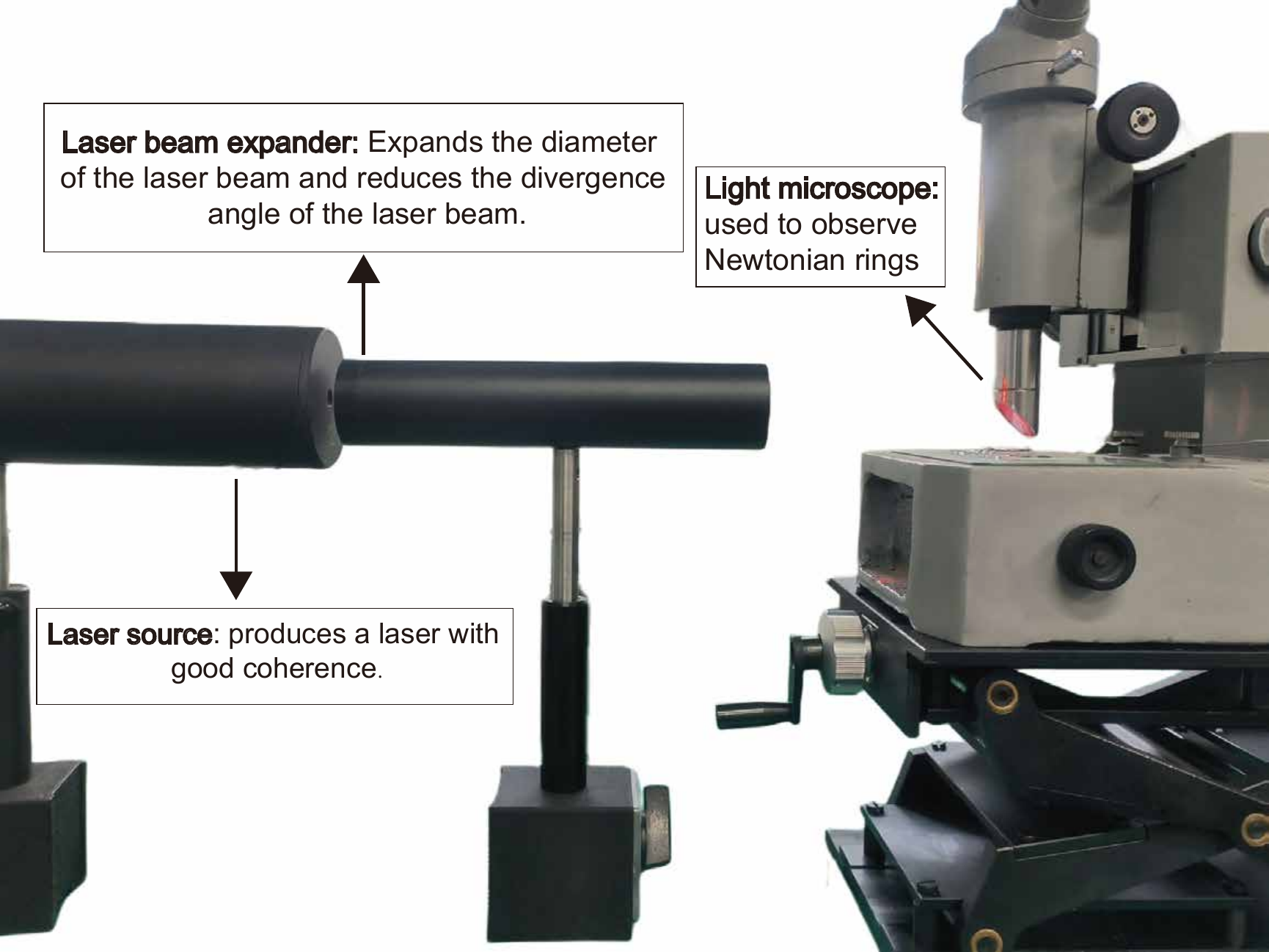}
\caption{\label{fig2} Physical drawing of the device, on which we labeled the laser source, the laser beam expander and the optical microscope.}
\end{figure}

\subsection{Operational procedures}

First add an appropriate amount of the liquid to be measured to the container and place the container under the microscope. After the liquid is stabilized, the optical path is adjusted as necessary until a clear interference pattern appears in the eyepiece. Subsequently, the scale on the eyepiece was calibrated according to the scale on the eyepiece and the interference pattern was photographed using a camera.

As shown in Fig. 3, in order to facilitate the subsequent data reading and measurement, after obtaining the interference pattern, we used the Robert operator in IS edge detection to extract and sharpen the edges of the image\cite{chaple2015comparisions}.The Robert operator can compensate away the noise present in the interference ring, thus enhancing the edges of the gray-scale jump region and other key regions, so as to make the contour of the interference ring clearer and the overall image effect optimized.

\begin{figure*}
\includegraphics[width=0.8\linewidth]{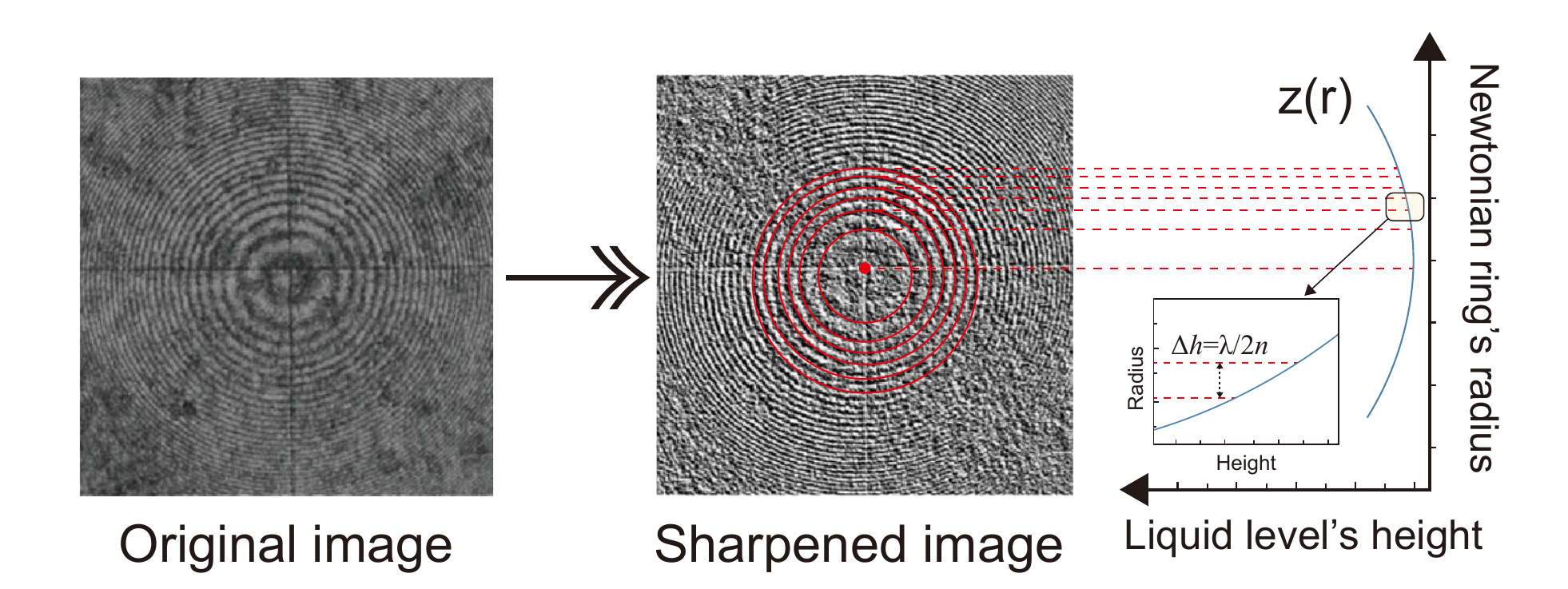}
\caption{\label{fig3}Example of interferometric ring image processing process, we will import the original image after sharpening into the computer, through the calibration of all levels of interferometric dark ring and the theoretical curve comparison to get the corresponding results of the inverse performance of the shape of the liquid surface.}
\end{figure*}

In order to reduce the error, it is necessary to pay attention to the following in the experiment: (1) the optical path and the container need to be kept strictly horizontal; (2) the interference pattern needs to be in the center of the field of view as much as possible, so that it is easy to read the interference fringes; (3) since the interference fringes have a certain width, the average of the inner and outer radius of the fringes will be taken as the fringes' radius; (4) the addition of samples should not be too much, so as to avoid that the height of the liquid exceeds the coherence length of the light source\cite{fowles1989introduction}; (5) The measurement time should not be too long to avoid the evaporation of the liquid to be measured or the small fluctuations of the ambient temperature have an effect on the measurement results.

\section{Results and discussion}

\subsection{Introduction of the translation factor C}

The surface tension coefficient of water was first measured at an ambient temperature of $25℃$. When the ambient temperature is $25℃$, the measured density of water is $1000.5 kg/m^3$. The reference value of the surface tension coefficient of water is $0.07201 N/m$. The measurement curve of the liquid surface shape is obtained by reading the radius of the equal-thickness interference fringes, and the measured liquid surface shape is fitted with (6), as shown in Fig. 4. After fitting, the surface tension coefficient of water at an ambient temperature of $25°C$ is $0.07696 N/m$, and the relative error with the reference value is $6.9\%$, which deviates from the reference value, and the reason for this is that the height of the zero-level interference fringes can not be determined at the time of measurement\cite{fowles1989introduction}. In order to solve the problem that the height of the zero-level interference fringes cannot be determined, we set the height of the first-level fringes as $0$, and introduce a translation factor in the fitting to make the curve $z(r)$ translates up and down by a certain distance, as shown in Fig. 4. $C$ can be derived from the fitting. After the translation, the surface tension coefficient of water is calculated to be $0.07266 N/m$, which is only $0.9\%$ relative error from the control value.

\begin{figure}
\includegraphics[width=\linewidth]{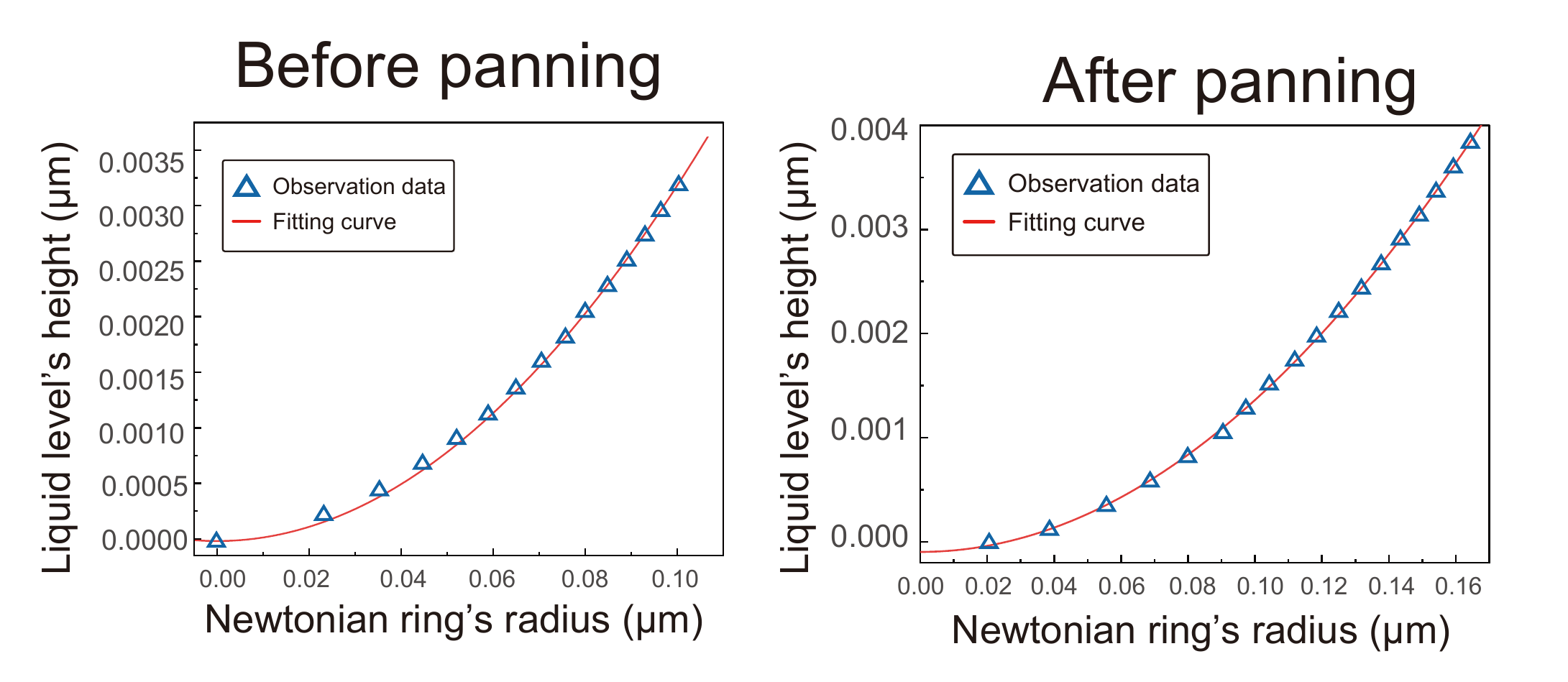}
\caption{\label{fig4}Fitting the measured values of the liquid level shape under the liquid level shape model.}
\end{figure}

\subsection{Measurement of surface tension coefficients of different concentrations of alcohol solutions}

To further validate the accuracy of the present method, we measured the surface tension coefficients of different concentrations of alcohol solutions. The surface tension coefficients of different concentrations of alcohol solutions were measured at temperature, different concentrations of alcohol solutions were configured using a pipette gun with an analytical balance, and the density of the liquid to be measured was measured , and the data are shown in Table 1.

\begin{table}
\caption{\label{tab:table2}Density of different concentrations of alcohol solutions.}
\begin{ruledtabular}
\begin{tabular}{cccccccc}
concentration (\%) & $0$ & $10$ & $20$ & $30$ & $40$ & $50$ & $60$\\
\hline
intensity $(kg/m^3)$ & $1000.5$ & $981.6$ & $968.2$ & $953.1$ & $935.3$ & $913.0$ & $891.2$\\
\end{tabular}
\end{ruledtabular}
\end{table}

The surface tension coefficients of different concentrations of alcoholic solutions were measured using the present method in $25℃$ environment, the surface tension coefficients of different concentrations of alcoholic solutions were measured using the present method and the relative errors to the reference values of literature\cite{sinzato2017experimental} are shown in Table 2.

\begin{table*}
\caption{\label{tab:table2}Measured values of surface tension coefficients for different concentrations of alcohol solutions.}
\begin{ruledtabular}
\begin{tabular}{cccc}
concentration (\%) & measured value $(N/m)$ & control values $(N/m)$ & $\frac{\sigma_{mea}-\sigma_{ref}}{\sigma_{ref}}$ $(\%)$ \\
\hline
$0$ & $0.07266$ & $0.07201$ & $0.9$ \\
$10$ & $0.04746$ & $0.04664$ & $1.8$ \\
$20$ & $0.03912$ & $0.04028$ & $-2.9$ \\
$30$ & $0.03679$ & $0.03767$ & $-2.3$ \\
$40$ & $0.03460$ & $0.03415$ & $1.3$ \\
$50$ & $0.03257$ & $0.03223$ & $1.1$ \\
$60$ & $0.03035$ & $0.03120$ & $-2.7$ \\
\end{tabular}
\end{ruledtabular}
\end{table*}

The traditional GUM\cite{wubbeler2020transferability} uncertainty assessment method based on the uncertainty propagation rate has significant limitations, which requires that the model of the measurement must be linear or linearizable, and the measured (output) is normally distributed or distribution, in our model, the data processing contains an optimization process\cite{harris2014monte}, so it is not possible to obtain the functional relationship between the data and the result directly, and the model of this nonlinear measurement cannot be used to calculate the uncertainty by the traditional GUM. Therefore we used Monte Carlo algorithm to calculate the uncertainty of the experiment. First simulate the data for the measured radius , which is in the range of $(r_i-\Delta r,r_i+\Delta r)$ the range of the Gaussian distribution of random numbers, take the $95\%$ confidence interval, to generate multiple sets of simulated data $\sigma$; then simulation calculation, that is, to calculate each set of simulated data value, the standard deviation of the simulation calculation results as the uncertainty of this experiment, the results are shown in Table 3.

\begin{table}
\caption{\label{tab:table3}Uncertainties of surface tension coefficients for different concentrations of alcohol solutions.}
\begin{ruledtabular}
\begin{tabular}{cccccccc}
concentration (\%) & $0$ & $10$ & $20$ & $30$ & $40$ & $50$ & $60$\\
\hline
intensity $(×10^{-5}N/m)$ & $44$ & $70$ & $62$ & $81$ & $52$ & $70$ & $63$\\
\end{tabular}
\end{ruledtabular}
\end{table}

In order to show the measurement results more intuitively, we plotted the variation of the surface tension coefficient of alcohol with concentration. As shown in Fig. 5, the measurement results of different concentrations of alcohol solutions using the present method are close to the reference value of the literature\cite{sinzato2017experimental}, and the maximum relative error with the reference value is not more than $3\%$, and the uncertainty is within a reasonable range, which can be regarded as accurate measurement. This indicates that in the range of $0-60\%$ alcohol concentration, accurate measurement can be realized by using this method.

\begin{figure}
\includegraphics[width=\linewidth]{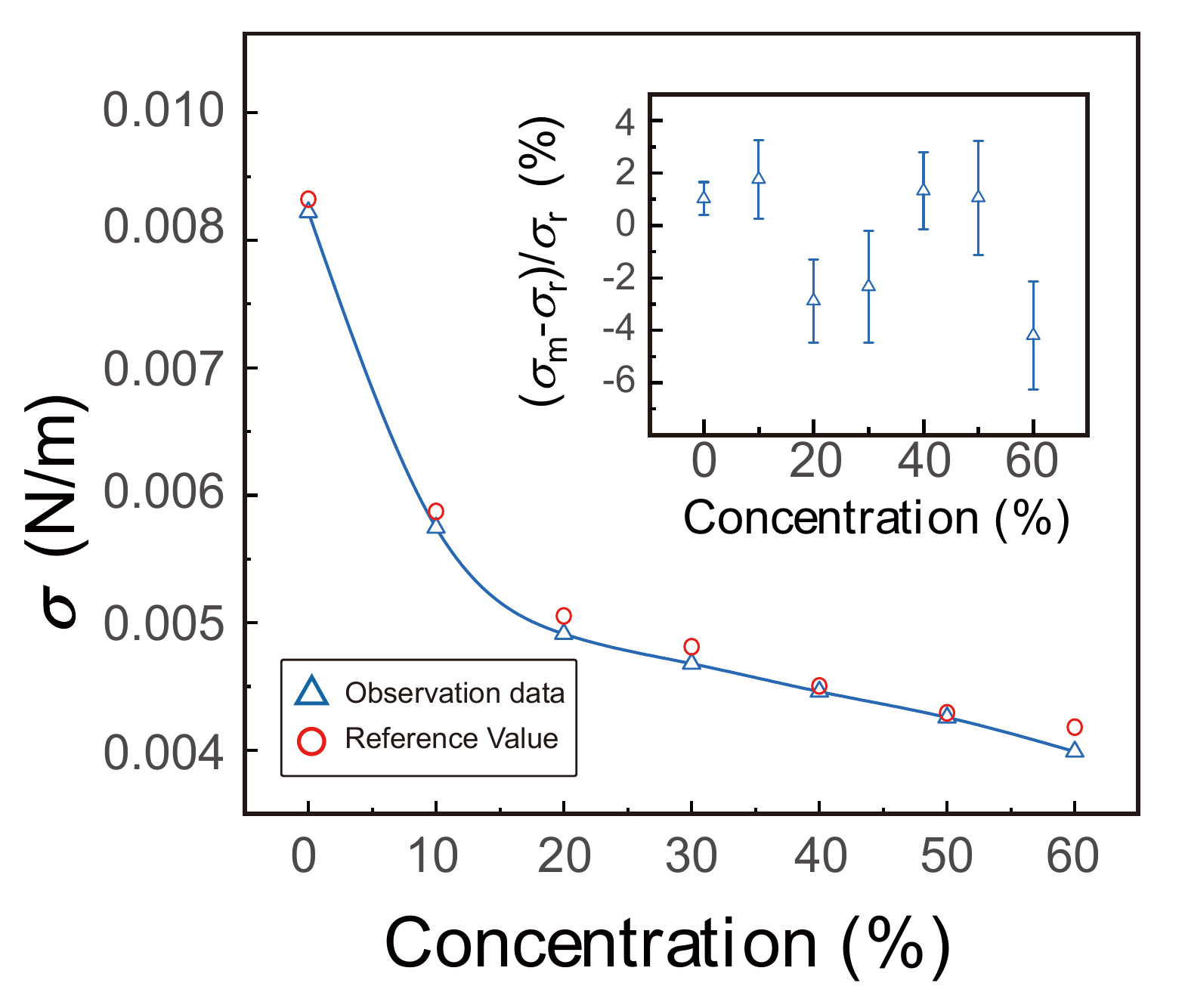}
\caption{\label{fig5}Plot of surface tension coefficient of alcohol solution as a function of concentration.}
\end{figure}

\subsection{Effect of containers on measurement results}

The shape of the liquid surface is related to the material and size of the container. The shape of the liquid surface formed when the liquid is in contact with different materials and containers with different radii may be different, but for the same liquid, the surface tension coefficient is only related to the composition of the liquid and the ambient temperature, and has nothing to do with the material of the container. In order to verify that different containers do not affect the measurement, the surface tension coefficient in different containers is measured using water as an example.

Measurement of the surface tension coefficient of water in containers of different diameters , surface tension coefficients of water in containers of different materials. The measurement results and relative errors are shown in Table 4.

\begin{table}
\caption{\label{tab:table4}Surface tension coefficients of water in containers of different materials.}
\begin{ruledtabular}
\begin{tabular}{ccc}
Container material & Measured value $(N/m)$ & $\frac{\sigma_{mea}-\sigma_{ref}}{\sigma_{ref}}$ $(\%)$\\
\hline
polytetrafluoroethylene & $0.07258$ & $0.8$ \\
acrylic (loanword) & $0.07120$ & $-1.1$ \\
fiberglass & $0.07334$ & $1.8$ \\
\end{tabular}
\end{ruledtabular}
\end{table}

The surface tension coefficients of water in containers of different diameters made of glass were measured. The measurement results and relative errors are shown in Table 5.

\begin{table}
\caption{\label{tab:table5}Surface tension coefficients of water in vessels of different diameters.}
\begin{ruledtabular}
\begin{tabular}{ccc}
Diameter $(cm)$ & Measured value $(N/m)$ & $\frac{\sigma_{mea}-\sigma_{ref}}{\sigma_{ref}}$ $(\%)$\\
\hline
$1.00$ & $0.07133$ & $-0.9$ \\
$1.50$ & $0.07292$ & $1.3$ \\
$1.80$ & $0.07271$ & $1.0$ \\
$2.00$ & $0.07145$ & $-0.8$ \\
\end{tabular}
\end{ruledtabular}
\end{table}

Using surface tension coefficients of water in vessels of different diameters the control variable method, measurements were made in a series of containers of different materials of the same diameter and a series of containers of different diameters of the same material. As can be seen from Tables 4 and 5, measured in different containers, the measured values are closer, and the relative error with the reference value are below $1.8\%$, indicating that the shape of the liquid surface in different materials and different sizes of containers is different, but it does not affect the measurement results.

\subsection{Measurement of liquid-liquid interfacial tension coefficients}

In addition to the liquid-gas surface tension coefficient, interfacial tension, i.e., liquid-liquid interfacial tension, also exists at the interface between two immiscible liquids. The liquid-liquid interfacial tension coefficient is similar in nature to the surface tension coefficient and plays an important role in the field of studying two-phase flow. Because the liquid-liquid interface is characterized by difficulty in observation and contact, the liquid-liquid interfacial tension coefficient is more difficult to measure than the surface tension coefficient. In this method, however, since the liquid-liquid interface can form a reflective surface, it can also be measured using this method. Different from the liquid-gas interface, here it is necessary to correct the The correction of Equation:

\begin{equation}
\sigma\left[\frac{rz_{rr}+z_r}{r(1+z_r)^{\frac{1}{2}}}-\frac{2}{R}\right]=\Delta \rho gz
\end{equation}

where $\Delta \rho$ is the difference in density between the lower and upper liquids, and since the density of the lower liquid is greater than that of the upper liquid, the value of $\Delta \rho$ constant positive value.

In order to prevent reflected light from the uppermost liquid-air interface from interfering with the experiments, it is necessary to add an excess of the upper liquid to beyond the coherence length of the light. We used the same method to measure the water-dodecane interfacial tension coefficient and the water-heptane interfacial tension coefficient, and the measured values and the errors relative to the reference values\cite{ghatee2014highly} are shown in Table 6. Analyzing the measurement data, it can be seen that under the same experimental conditions as in the literature, the relative error between the measured water-dodecane interfacial tension coefficient and water-heptane interfacial tension coefficient and the reference value is less than $3.1\%$, which is of high measurement accuracy and provides a simple and accurate method for the measurement of liquid-liquid interfacial tension coefficient.

\begin{table*}
\caption{\label{tab:table6}Measurement of liquid-liquid interfacial tension coefficients.}
\begin{ruledtabular}
\begin{tabular}{cccc}
liquid-liquid interface & Measured value $(N/m)$ & Reference value $(N/m)$ &$\frac{\sigma_{mea}-\sigma_{ref}}{\sigma_{ref}}$ $(\%)$\\
\hline
Water-dodecane & $0.05130$ & $0.05270$ & $-2.7$ \\
Water-heptane & $0.05185$ & $0.05030$ & $3.1$
\end{tabular}
\end{ruledtabular}
\end{table*}

\section{Conclusion}

In this paper, we propose a new method to measure the surface tension coefficient based on equal-thickness interference, derive a theoretical model of the axisymmetric liquid surface shape, measure the liquid surface shape by equal-thickness interference, and obtain the surface tension coefficient of the liquid by fitting the measured data of the liquid surface shape with the derived model of the liquid surface shape, so as to realize the surface tension coefficient in a non-contact manner. In addition we introduce a translation factor based on the traditional Newton's ring experiment which allows the curve up and down translation, which effectively solves the problem of uncertainty in the position of the zero-level stripes in the traditional Newton's ring experiment, and reduces the relative error from $6.9\%$ to $0.9\%$.

In this paper, the surface tension coefficients of alcohol solutions of different concentrations were measured, and the control variable method was utilized to verify that the measurement results were independent of the size and material of the containers, and the designed measurement method was applied to the measurement of the tension coefficients of liquid-liquid interfaces. In the measurement of surface tension coefficients of different concentrations of alcohol solutions, the relative error of the same reference value does not exceed $2.9\%$; in the measurement of liquid-liquid interfacial tension coefficients of different incompatible liquids, the relative error of the same reference value does not exceed $3.1\%$. The experimental results show that the measurement method of surface tension coefficient based on isothermal interference is rigorous and accurate, and the measurement device and operation procedure are relatively simple, in addition, the method expands the measurement range and provides a simple and accurate measurement method for the difficult liquid-liquid interfacial tension coefficient. This method broadens the idea of surface tension coefficient measurement, and is also of great significance to the study of the evolution of multiphase interfacial dynamics in fluid dynamics.

\nocite{*}
\bibliography{aipsamp}

\end{document}